\documentclass{article}

\usepackage[mathletters]{ucs}
\usepackage[utf8x]{inputenc}
\usepackage{here}
\usepackage{url}
\usepackage{syntax}
\usepackage{amssymb}
\usepackage{amsmath}

\begin{document}

\title{Scalar and Tensor Parameters for Importing Tensor Index Notation including Einstein Summation Notation}
\author{Satoshi Egi\\
  Rakuten Institute of Technology}

\maketitle

\begin{abstract}
In this paper, we propose a method for importing tensor index notation, including Einstein summation notation, into functional programming.
This method involves introducing two types of parameters, i.e, scalar and tensor parameters, and simplified tensor index rules that do not handle expressions that are valid only for the Cartesian coordinate system, in which the index can move up and down freely.
An example of such an expression is ``$c = A_i B_i$''.
As an ordinary function, when a tensor parameter obtains a tensor as an argument, the function treats the tensor argument as a whole.
In contrast, when a scalar parameter obtains a tensor as an argument, the function is applied to each component of the tensor.
In this paper, we show that introducing these two types of parameters and our simplified index rules enables us to apply arbitrary user-defined functions to tensor arguments using index notation including Einstein summation notation without requiring an additional description to enable each function to handle tensors.
\end{abstract}

\section{Introduction}\label{intro}

Tensor analysis is one of the fields of mathematics in which we can easily find notations that have not been imported into popular programming languages~\cite{fleisch2011student,schutz1980geometrical}.
Index notation is one such notation widely used by mathematicians to describe expressions in tensor analysis concisely.
This paper proposes a method for importing it into programming.

Tensor analysis is also a field with a wide range of application.
For example, the general theory of relativity is formulated in terms of tensor analysis.
In addition, tensor analysis plays an important role in other theories in physics, such as fluid dynamics.
In fields more familiar to computer scientists, tensor analysis is necessary for computer vision~\cite{hartley2003multiple}.
Tensor analysis also appears in the theory of machine learning to handle multidimensional data.
The importance of tensor calculation is increasing day by day even in computer science.

Concise notation for tensor calculation in programming will simplify technical programming in many areas.
Therefore, it is important to develop a method for describing tensor calculation concisely in programs.


The novelty of this paper is that it introduces two types of parameters, \textit{tensor parameters} and \textit{scalar parameters}, to import tensor index notation and enables us to apply arbitrary user-defined functions to tensor arguments using index notation without requiring an additional description to enable each function to handle tensors.
Tensor and scalar parameters are used to define two types of functions, \textit{tensor functions} and \textit{scalar functions}, respectively.
Tensor functions are functions that involve contraction of the tensors provided as an argument, and scalar functions are not.
For example, the inner product of vectors and the multiplication of matrices are tensor functions, because they involve contraction of the tensors.
Most other functions such as ``\texttt{+}'', ``\texttt{-}'', ``\texttt{*}'', and ``\texttt{/}'' are scalar functions.
We also simplified tensor index rules by removing a rule to handle the expressions such as ``$c = A_i B_i$'' that are valid only for the Cartesian coordinate system, in which the index can move up and down freely.
It enables us to use the same index rules for all user-defined functions.
We will discuss that in Sections~\ref{related} and~\ref{egison}.

The method proposed in this paper has already been implemented in the Egison programming language~\cite{egison}.
Egison is a functional language that has a lazy evaluation strategy and supports symbolic computation.
In reading this paper, one can think of that language as an extended Scheme to support symbolic computation.

Figure~\ref{fig:minFunction} shows the definition of the \texttt{min} function as an example of a scalar function.
The \texttt{min} function takes two numbers as arguments and returns the smaller one.
``\texttt{\$}'' is prepended to the beginning of the parameters of the \texttt{min} function.
It means the parameters of the \texttt{min} function are scalar parameters.
When a scalar parameter obtains a tensor as an argument, the function is applied to each component of the tensor as Figures~\ref{fig:minDiff} and~\ref{fig:minSame}.
As Figure~\ref{fig:minDiff}, if the indices of the tensors of the arguments are different, it returns the tensor product using the scalar function as the operator.
As Figure~\ref{fig:minSame}, if the indices of the tensors given as arguments are identical, the scalar function is applied to each corresponding component.
Thus the \texttt{min} function can handle tensors even though it is defined without considering tensors.
The function name ``\texttt{\$min}'' is also prefixed by ``\texttt{\$}'', but just as a convention of Egison.
Thus it can be ignored.

\begin{figure}[t]
  \begin{center}
  {\footnotesize
\verb|(define $min (lambda [$x $y] (if (less-than? x y) x y)))|
  }
  \end{center}
  \caption{Definition of the \texttt{min} function}
  \label{fig:minFunction}
  \medskip
  {\small
    $min(\begin{pmatrix} 1 \\ 2 \\ 3 \\ \end{pmatrix}_{i},  \begin{pmatrix} 10 \\ 20 \\ 30 \\ \end{pmatrix}_{j}) = \begin{pmatrix} min(1,10) & min(1,20) & min(1,30) \\ min(2,10) & min(2,20) & min(2,30) \\ min(3,10) & min(3,20) & min(3,30) \\ \end{pmatrix}_{ij}
    = \begin{pmatrix} 1 & 1 & 1 \\ 2 & 2 & 2 \\ 3 & 3 & 3 \\ \end{pmatrix}_{ij}$
    }
  \caption{Application of the \texttt{min} function to the vectors with different indices}
  \label{fig:minDiff}
  \medskip
  {\small
    $min(\begin{pmatrix} 1 \\ 2 \\ 3 \\ \end{pmatrix}_{i},  \begin{pmatrix} 10 \\ 20 \\ 30 \\ \end{pmatrix}_{i}) = \begin{pmatrix} min(1,10) & min(1,20) & min(1,30) \\ min(2,10) & min(2,20) & min(2,30) \\ min(3,10) & min(3,20) & min(3,30) \\ \end{pmatrix}_{ii}
  = \begin{pmatrix} min(1,10) \\ min(2,20) \\ min(3,30) \end{pmatrix}_{i} = \begin{pmatrix} 1 \\ 2 \\ 3 \end{pmatrix}_{i}$ 
    }
  \caption{Application of the \texttt{min} function to the vectors with identical indices}
  \label{fig:minSame}
\end{figure}

Figure~\ref{fig:dotFunction} shows the definition of the ``\texttt{.}'' function as an example of a tensor function.
``\texttt{.}'' is a function for multiplying tensors.
``\texttt{\%}'' is prepended to the beginning of the parameters of the ``\texttt{.}'' function.
It means the parameters of the ``\texttt{.}'' function are tensor parameters.
As with ordinary functions, when a tensor is provided to a tensor parameter, the function treats the tensor argument as a whole.
When a tensor with indices is provided, it is passed to the tensor function maintaining its indices.

In Figure~\ref{fig:dotFunction}, ``\texttt{+}'' and ``\texttt{*}'' are scalar functions for addition and multiplication, respectively.
\texttt{contract} is a primitive function to contract a tensor that has pairs of a superscript and subscript with identical symbols.
We will explain the semantics of \texttt{contract} expression in Section~\ref{egisonIndex}.
Figure~\ref{fig:dotSame} shows the example for calculating the inner product of two vectors using the ``\texttt{.}'' function.
We can use the ``\texttt{.}'' function for any kind of tensor multiplication such as tensor product and matrix multiplication as well as inner product.

\begin{figure}[t]
  {\footnotesize
\verb|(define $. (lambda [%t1 %t2] (contract + (* t1 t2))))|
  }
  \caption{Definition of the ``\texttt{.}'' function}
  \label{fig:dotFunction}
  \medskip
  {\small
  $\begin{pmatrix} 1 \\ 2 \\ 3 \\ \end{pmatrix}^{i} \cdot \begin{pmatrix} 10 \\ 20 \\ 30 \\ \end{pmatrix}_{i} = contract(+, \begin{pmatrix} 10 & 20 & 30 \\ 20 & 40 & 60 \\ 30 & 60 & 90 \\ \end{pmatrix}^{i}_{\;i})
  = 10 + 40 + 90 = 140$
  \\
  $\begin{pmatrix} 1 \\ 2 \\ 3 \\ \end{pmatrix}_{i} \cdot \begin{pmatrix} 10 \\ 20 \\ 30 \\ \end{pmatrix}_{i} = contract(+, \begin{pmatrix} 10 \\ 40 \\ 90 \\ \end{pmatrix}_{i}) = \begin{pmatrix} 10 \\ 40 \\ 90 \\ \end{pmatrix}_{i}$
  \\
  $\begin{pmatrix} 1 \\ 2 \\ 3 \\ \end{pmatrix}_{i} \cdot \begin{pmatrix} 10 \\ 20 \\ 30 \\ \end{pmatrix}_{j} = contract(+, \begin{pmatrix} 10 & 20 & 30 \\ 20 & 40 & 60 \\ 30 & 60 & 90 \\ \end{pmatrix}_{ij})
  = \begin{pmatrix} 10 & 20 & 30 \\ 20 & 40 & 60 \\ 30 & 60 & 90 \\ \end{pmatrix}_{ij}$ \\
  }
  \caption{Application of the ``\texttt{.}'' function}
  \label{fig:dotSame}
\end{figure}

Here we introduce a more mathematical example.
The expression in Figure~\ref{fig:inMathShort} from tensor analysis can be expressed in Egison as shown in Figure~\ref{fig:inEgisonShort}.
When the same mathematical expression is expressed in a general way in the Wolfram language, it becomes a program such as the one shown in Figure~\ref{fig:inWolframShort}.
In the Wolfram language, it is assumed that all dimensions corresponding to each index of the tensor are a constant ``\texttt{M}''.

\begin{figure}[t]
\[R^{i}_{\;jkl} = \frac{\partial \Gamma^{i}_{\;jl}}{\partial x^k} - \frac{\partial \Gamma^{i}_{\;jk}}{\partial x^l} + \Gamma^{m}_{\;jl} \Gamma^{i}_{\;mk} - \Gamma^{m}_{\;jk} \Gamma^{i}_{\;ml} \]
  \caption{Formula of Riemann curvature tensor}
  \label{fig:inMathShort}
  \medskip
{\footnotesize
\begin{verbatim}
(define $R~i_j_k_l
  (with-symbols {m}
    (+ (- (∂/∂ Γ~i_j_l x~k) (∂/∂ Γ~i_j_k x~l))
       (- (. Γ~m_j_l Γ~i_m_k) (. Γ~m_j_k Γ~i_m_l)))))
\end{verbatim}
}
  \caption{Egison program that represents the formula in Figure~\ref{fig:inMathShort}}
  \label{fig:inEgisonShort}
  \medskip
{\footnotesize
\begin{verbatim}
R=Table[D[Γ[[i,j,l]],x[[k]]] - D[Γ[[i,j,k]],x[[l]]]
       +Sum[Γ[[m,j,l]] Γ[[i,m,k]]
          - Γ[[m,j,k]] Γ[[i,m,l]],
            {m,M}],
        {i,M},{j,M},{k,M},{l,M}]
\end{verbatim}
}
  \caption{Wolfram program that represents the formula in Figure~\ref{fig:inMathShort}}
  \label{fig:inWolframShort}
\end{figure}

Note that a double loop consisting of the \texttt{Table} and \texttt{Sum} expressions appears in the program in the Wolfram language, whereas the program in Egison is flat, similarly to the mathematical expression.
This is achieved by using tensor index notation in the program.
In particular, the reason that the loop structure by the \texttt{Sum} expression in the Wolfram language does not appear in the Egison expression to express $\Gamma^{m}_{\;jk} \Gamma^{i}_{\;ml} - \Gamma^{m}_{\;jl} \Gamma^{i}_{\;mk}$ is that the ``\texttt{.}'' function in Egison can handle Einstein summation notation.

The part that we would like the reader to pay particular attention to in this example is the Egison program ``\verb|(∂/∂ Γ~i_j_k x~l)|'' expressing $\frac{\partial \Gamma^{i}_{\;jk}}{\partial x^l}$ in the first term on the right-hand side.
In the Wolfram language, the differential function ``\texttt{D}'' is applied to each tensor component, but the differential function ``\texttt{∂/∂}'' is applied directly to the tensors in Egison.

The differential function ``\texttt{∂/∂}'' is defined in an Egison program as a scalar function.
When a tensor is provided as an argument to a scalar function, the function is applied automatically to each component of the tensor.
Therefore, when defining a scalar function, it is sufficient to consider only a scalar as its argument.
That is, in the definition of the ``\texttt{∂/∂}'' function, the programmer need only write the program for the case in which the argument is a scalar value.
Despite that, the program ``\verb|(∂/∂ Γ~i_j_k x~l)|'' returns a fourth-order tensor.

The program ``\verb|(∂/∂ Γ~i_j_k x~l)|'' returns a fourth-order tensor with superscript ``\texttt{i}'', subscript ``\texttt{j}'', subscript ``\texttt{k}'', and subscript ``\texttt{l}'' from left to right.
Here the superscript ``\verb|~l|'' of ``\verb|x~l|'' is inverted and becomes the subscript ``\verb|_l|'', because the differential operator is a special function in tensor analysis and the indices of the tensor applied as the denominator of $\frac{\partial}{\partial}$ are inverted upside down.
We will discuss that in detail in Section~\ref{egison}.

Thus, we can naturally import tensor index notation including Einstein notation into programming if we clearly distinguish between tensor functions such as ``\texttt{.}'' and scalar functions such as ``\texttt{+}'' and ``\texttt{∂/∂}''.
This paper explains this.

The remainder of this paper is structured as follows.
In Section~\ref{related}, we explain existing work to import index notation into programming and its problems.
In Section~\ref{egison}, we describe our new method for importing index notation and explain how it solves the existing problems.
In Section~\ref{demo}, we present a program for calculating the Riemannian curvature tensor in the Wolfram language and Egison, and we show how the expression is simplified by the method described in Section~\ref{egison}.
In the final section, we summarize the contribution of this paper and future issues.

\section{Existing Work on Index Notation}\label{related}

There are two existing methods for using index notation in programming, a method that introduces special operators supporting index notation and a method that introduces special syntax for index notation.

Using the first method enables index notation to be represented directly in a program.
However, this has the disadvantage that index notation can be used only by functions that are specially prepared to use it.

In the second method, we describe the computation of the tensor using syntax such as the \texttt{Table} expression of the Wolfram language~(Figure \ref{fig:inWolframShort}).
This method has the advantage that we can use an arbitrary function defined for scalar values also for tensor operations, similarly to the differential function ``\texttt{D}'' in Figure~\ref{fig:inWolframShort}.
However, this method has the disadvantage that we cannot directly apply user-defined functions to tensor arguments using index notation.
As the result, the description of a program becomes more complicated than the description directly using index notation, as we explained in the previous section using Figure~\ref{fig:inEgisonShort} and Figure~\ref{fig:inWolframShort}.


\subsection{Introduction of Index Notation by Special Operators}\label{maxima}

For existing work using this method, there is Maxima~\cite{maximaWeb,toth2005tensor}, a computer algebra system that introduces index notation through the extension library itensor, as well as Ahalander's work~\cite{aahlander2002einstein}, which implements index notation on C++.
These studies introduce index notation by implementing two special functions ``$+$'' and ``$\cdot$'' that support tensor index notation.

``$+$'' is a function that sums the components of two tensors given as arguments.
``$\cdot$'' is a function that takes the tensor product of the two tensors given as arguments and takes the sum of the trace if there are pairs of the superscript and subscript with the same index variable.

``$+$'' and ``$\cdot$'' take different actions on tensors with the same index combination.
For example, in the following expression, programs corresponding to $(1)$, $(4)$, and $(5)$ are invalid.
It is natural to implement index notation in this way because these expressions are meaningless in mathematics.

$(1) \qquad c_{ij} = a_{i} + b_{j}$

$(2) \qquad c_{ij} = a_{i} \cdot b_{j}$

$(3) \qquad c_{ij} = a_{ij} + b_{ij}$

$(4) \qquad c_{ij} = a_{ij} \cdot b_{ij}$

$(5) \qquad c = a^{ij} + b_{ij}$

$(6) \qquad c = a^{ij} \cdot b_{ij}$

Especially, Ahalander's work ~\cite{aahlander2002einstein} can interpret the following expressions, which are meaningless when dealing with a general coordinate system other than the Cartesian coordinate system.
In Ahalander's work~\cite{aahlander2002einstein}, $(7)$ is interpreted as equivalent to $(3)$, and $(8)$ is interpreted as equivalent to $(6)$.
In the range dealing with the Cartesian coordinate system, the index can move up and down freely, with the result that $(7)$ and $(3)$, and $(8)$ and $(6)$ are equivalent in mathematics, making such an interpretation useful.

$(7) \qquad c_{ij} = a^{ij} + b_{ij}$

$(8) \qquad c = a_{ij} \cdot b_{ij}$

Thus, in these work, the index rules of ``$+$'' and ``$\cdot$'' for tensors are defined separately.
As a result, ``$+$'' and ``$\cdot$'' for tensors in the existing work are special operators prepared in the library for index notation, with the consequence that we need to edit the library directly to add new functions that support index notation.
That is, it takes substantial effort to define the original operators for tensors.

There are important operations in tensor analysis other than simply adding and multiplying two tensors.
For example, in tensor analysis, as we see in connection with ``\verb|(∂/∂ Γ~i_j_k x~l)|'' in Figure~\ref{fig:inEgisonShort}, we often differentiate the components of a tensor with respect to the components of another tensor.
It is a serious problem that programmers cannot add such operations easily.



\subsection{Introduction of Tensor Index Notation by Special Syntax}\label{wolfram}

The crucial difference of this method from the method in Section~\ref{maxima} is that we no longer consider the index rules for ``$+$'' and ``$\cdot$'' separately.
In this method, we regard ``$+$'' for tensors as a function that computes the tensor product using ``$+$'' for scalars, and the function for computing the sum of the trace for contraction is undefined, as shown in Figures~\ref{fig:tabFunc}.
Then we can regard ``$+$'' as a variation of ``$\cdot$'' that never contract the result of the tensor product. 
As shown Figure~\ref{fig:inPlusIndex}, we can interpret ``$+$'' correctly even if we regard ``$+$'' in the above manner.
This idea simplifies the index rules.
In this method, we control the way we combine tensors only through their indices.

\begin{figure}
    \begin{tabular}{ | l | l | l |}
      \hline
      & operator for tensor product & operator for contraction \\ \hline
      $\cdot$ & * & $+$ \\ \hline
      $+$ & $+$ & undefined \\ \hline
      $*$ & $*$ & undefined \\ \hline
      ∂/∂ & ∂/∂ & undefined \\ \hline
    \end{tabular}
  \caption{All scalar functions are regarded as variations of the ``$\cdot$'' function whose operator for tensor product is itself and one for contraction is undefined.}
  \label{fig:tabFunc}
  \medskip
  $\begin{pmatrix} 1 \\ 2 \\ 3 \\ \end{pmatrix}_{i} + \begin{pmatrix} 10 \\ 20 \\ 30 \\ \end{pmatrix}_{i} = \begin{pmatrix} 11 & 21 & 31 \\ 12 & 22 & 32 \\ 13 & 23 & 33 \\ \end{pmatrix}_{ii} = \begin{pmatrix} 11 \\ 22 \\ 33 \\ \end{pmatrix}_{i}$
  \caption{Interpretation of ``$+$'' as an operator for tensor product}
  \label{fig:inPlusIndex}
\end{figure}

Even if this idea is introduced, the expressions $(2)$, $(3)$, and $(6)$ are still interpreted correctly as before.
However, the expressions $(1)$ and $(4)$, which are invalid in mathematics, do not cause errors in this interpretation.
Expression $(5)$ still causes an error since the function for calculating the sum of the trace to contract the tensor is undefined for ``$+$''.
Thus, the introduction of this idea renders the interpretation of some expressions meaningless in mathematics, but we had assigned a higher priority to making the index rules more concise.

This method is adopted primarily in the Wolfram Language, which has a tensor generation syntax, the Table expression~\cite{wolframTable}.
In the Wolfram language, the \texttt{Table} expression plays a major role in describing operations that deal with tensors.
Using this syntax, we can use arbitrary functions to deal with tensors using a notation with similar to that of index notation.
Since this method has such advantages, a program using this method has been introduced by mathematicians in actual research.~\cite{maeda2016geometry,maeda2010program}

For example, the expression $A_{ij} + B_{ij}$ is expressed as follows using the \texttt{Table} expression.
In the following examples, it is assumed that all dimensions corresponding to each index of the tensors are a constant ``\texttt{M}''.

{\footnotesize
\begin{verbatim}
Table[A[[i, j]] + B[[i,j]],
      {i, M},{j, M}]
\end{verbatim}
}

The expression $A_{ij} B_{kl}$ is expressed as follows.

{\footnotesize
\begin{verbatim}
Table[A[[i, j]] * B[[k,l]],
      {i, M},{j, M},{k, M},{l, M}]
\end{verbatim}
}

In the Wolfram language, it is possible to express both addition of tensors and the tensor product in a unified manner using the \texttt{Table} expression.
In addition, we can use arbitrary functions as an operator for tensor product, similarly to the manner in which the differential function ``\texttt{D}'' is used in Figure~\ref{fig:inWolframShort}.
Thus, parameterization of an operator for tensor product, as described in Section~\ref{maxima} is realized in the Wolfram language.

A program that contracts tensors is described by combining the \texttt{Table} and \texttt{Sum} expressions.
For example, the expression $A^{ij} T_{jkl}$ is as follows.
In the Wolfram language, we do not explicitly specify whether tensor indices are superscripts or subscripts.
We need to determine from the context whether the index of the tensor in a program is a superscript or subscript.

{\footnotesize
\begin{verbatim}
Table[Sum[A[[i, j]] * T[[j, k, l]], {j, M}],
      {i, M},{k, M},{l, M}]
\end{verbatim}
}

We can use a different aggregate function instead of the \texttt{Sum} expression.
This means that the parameterization of an operator for contraction described in Section~\ref{maxima} is also realized in the Wolfram language.

In the Wolfram language, parameterizations of an operator for tensor product and contraction are both realized.
However, it has a disadvantage that it always requires to use the \texttt{Table} expression for tensor operation using index notation.
The Wolfram language does not allow to directly apply a function using index notation as Figures~\ref{fig:minDiff},~\ref{fig:minSame}, and~\ref{fig:dotSame} in Section~\ref{intro}.
Furthermore, it is impossible to modularize an operation, such as ``$\cdot$'' described in Section~\ref{maxima}, whose behavior varies depending on the combination of indices of the tensors of the argument.
That is, expressing such operations requires us to write the \texttt{Sum} expression nested in the \texttt{Table} expression every time, because the tensor of the Wolfram language does not contain index information.
Consequently, the program becomes more complicated than the mathematical expression.

Domain Specific Languages (DSLs) for computational chemistry~\cite{solomonik2015sparse} also support index notation including Einstein summarize notation in this method.
For example, the expression $A_{ij} T_{jkl}$ is expressed as follows.
In this work, superscripts and subscripts are not distinguished as the Wolfram language.

{\footnotesize
\begin{verbatim}
B["ikl"] = A["ij"] * T["jkl"]
\end{verbatim}
}

The above program directly represents index notation.
However, this method has two drawbacks.

First, it has a constraint that we always need to specify the type of the tensor returned by expressions using index notation.
That is, we need to write a left-side expression as ``\texttt{B["ikl"]}'' in the above example.
This constraint is caused by the ambiguity of the interpretation of the right-side expression of $(4)$ in Section~\ref{maxima}.
This work allows to interpret the right-side expression of $(4)$ in both manners, $(4)$ and $(8)$.
Therefore, we need to specify the manner in which we interpret the expressions using index notation by specifying the type of the tensor returned by them.

Our proposal eliminates this ambiguity by distinguishing superscripts and subscripts, and ceasing the interpretation of the expressions such as $(8)$ that are valid only for the Cartesian coordinate system in which index can move up and down freely.
It enables us to directly apply a function to tensor arguments using index notation as Figures~\ref{fig:minDiff},~\ref{fig:minSame}, and~\ref{fig:dotSame} in Section~\ref{intro}.

The second disadvantage is that it requires additional description to temporarily change a function for calculating the tensor product and trace.
In this work, the additive and multiplicative operators defined as the class methods of the class to which the elements of the tensor belong are used to caluculate the trace and tensor product, respectively.
This is the reason why the above program do not specify the operator to use when calculating the trace corresponding to the \texttt{Sum} expression in the above Wolfram program.
However, it requires additional description when we specify a function to use when calculating the tensor product as follows.
In the following example, we specify the \texttt{min} function as the operator for calculating the tensor product.

{\footnotesize
\begin{verbatim}
((Transform<>)([] (double a, double b, double & c){ c = min(a,b); }))(A["i"],B["j"],C["ij"]);
\end{verbatim}
}

The above program is equivalent to the following program.

{\footnotesize
\begin{verbatim}
for (int i=0; i<n; i++){
  for (int j=0; j<n; j++){
    C[i,j] = min(A[i],B[j]);
  }
}
\end{verbatim}
}

We avoid this problem by introducing the concept of two types of functions, scalar and tensor functions.
As the result, we can directly apply arbitrary user-defined functions using index notation, as we directly apply the ``\texttt{+}'', ``\texttt{-}'', ``\texttt{∂/∂}'', and ``\texttt{.}'' functions to tensors in Figure~\ref{fig:inEgisonShort} in Section~\ref{intro}.

\section{A New Method for Importing Index Notation into Programming}\label{egison}

As mentioned in Section~\ref{related}, the existing methods for importing tensor index notation have a disadvantage that we cannot directly apply arbitrary user-defined functions to tensor arguments using index notation.
We can overcome this disadvantage if we satisfy all of the following three conditions at the same time.

\begin{description}
 \item[No ambiguity in tensor index rules] We do not need to specify the tensor type of the return value of an expression using index notation.
 \item[Parameterization of an operator for tensor product] We can specify a function to use when calculating the tensor product with a parameter.
 \item[Parameterization of an operator for contraction] We can specify a function to use when calculating the sum of the trace to contract tensors with a parameter.
\end{description}

In the work explained in section~\ref{maxima}, the first condition is satisfied, but the second and third are not.
In contrast, in the work explained in Section~\ref{wolfram}, the second and third conditions are satisfied, but the first is not.
Therefore, both of the sections have relevant disadvantages.

This section discusses a means of satisfying all of these conditions simultaneously in programming languages.

\subsection{Grammar}\label{tensor-grammar}

Figure~\ref{fig:egisonGrammar} shows the syntax added to implement the proposed method in Egison.
By implementing the same syntax, we can import index notation into programming languages other than Egison.
In Section~\ref{egison}, we explain how to implement our proposal by explaining the semantics of this grammar.

\begin{figure*}
  \begin{center}
\begin{grammar}
  <expr> ::= <tensor> | `(' <expr> [<expr> ...] `)' | `(with-symbols {' [<symbol> ...] `}' <expr> `)' | `(tensor-map' <function> <tensor> `)' | `(contract' <function> <tensor> `)' | `(flip-indices' <tensor> `)'

  <tensor> ::= <tensor-data> [<index> ...]

  <tensor-data> ::= <variable-name> | <scalar> | <function> | `[|' <tensor-data> ... `|]' | `(generate-tensor ' <function> `{' [<natural-number> ...] `})'
  
  <index> ::= <index-type> <natural-number> | <index-type> <symbol> | <index-type> `#'

  <index-type> ::= `~' | `_' | `~_'

  <function> ::= `(lambda' `[' [<parameter> ...] `]' <expr> `)' | <builtin-scalar-function>

  <parameter> ::= `\%' <variable-name> | `$' <variable-name> | `*$' <variable-name>
  
  <builtin-scalar-function> ::= `+' | `-' | `*' | `/' | ...
\end{grammar}
  \end{center}
  \caption{Grammar for index notation}
  \label{fig:egisonGrammar}
\end{figure*}

In Figure~\ref{fig:egisonGrammar}, \synt{scalar} represents scalar values such as numbers (``\texttt{1}'', ``\texttt{2}'', ``\texttt{(/ 3 2)}'') and expressions (``\texttt{(+ x y)}'', ``\verb|x^2|'', ``\texttt{(cos θ)}'').
\texttt{tensor-map}, \texttt{contract}, \texttt{flip-indices}, and \texttt{generate-tensor} are primitive syntax to handle tensors.

We use Section~\ref{egisonIndex} and~\ref{sfTf} to explain our index reduction rules for tensors with symbolic indices.
We show various examples to show their validity in the subsequent sections.

\subsection{Reduction Rules for Tensors with Indices}\label{egisonIndex}


In this section, we show the index reduction rules for a single tensor with indices.

First, we explain the notation for expressing tensors in Egison.
Egison expresses a tensor by enclosing its components with ``\texttt{[|}'' and ``\texttt{|]}''.
We express a higher-order tensor by nesting this description, as we do for an $n$-dimensional array.

To access the components of a tensor, we add indices to the tensor.
Subscripts are represented by ``\verb|_|'' followed by a natural number after the tensor.
An arbitrary number of indices can be added to one tensor, though adding a number of indices larger than the rank of the target tensor results in an error.

{\footnotesize
\begin{verbatim}
[|[|11 12 13|] [|21 22 23|] [|31 32 33|]|]_2
;[|21 22 23|]

[|[|11 12 13|] [|21 22 23|] [|31 32 33|]|]_2_1
;21
\end{verbatim}
}


We can use symbols as well as natural numbers as indices.
Egison is a computer algebra system and supports symbolic computation.
Unbound variables are treated as symbols.
We declare the indices of a tensor by using the symbols for the indices.
If multiple indices of the same symbol appear, Egison converts it to the tensor composed of diagonal components for these indices.
After this conversion, the leftmost index symbol remains.
For example, the indices ``\verb|_i_j_i|'' convert to ``\verb|_i_j|''.

{\footnotesize
\begin{verbatim}
[|[|11 12 13|] [|21 22 23|] [|31 32 33|]|]_i_j
;[|[|11 12 13|] [|21 22 23|] [|31 32 33|]|]_i_j

[|[|11 12 13|] [|21 22 23|] [|31 32 33|]|]_i_i
;[|11 22 33|]_i

[|[|[|1 2|] [|3 4|]|] [|[|5 6|] [|7 8|]|]|]_i_j_i
;[|[|1 3|] [|6 8|]|]_i_j
\end{verbatim}
}

When three or more subscripts of the same symbol appear, Egison converts it to the tensor composed of diagonal components for all these indices.

{\footnotesize
\begin{verbatim}
[|[|[|1 2|] [|3 4|]|] [|[|5 6|] [|7 8|]|]|]_i_i_i
;[|1 8|]_i
\end{verbatim}
}

Egison supports two types of indices, both superscripts and subscripts.
A subscript is represented by ``\verb|_|''.
A superscript is represented by ``\texttt{\~}''.

Superscripts and subscripts behave symmetrically.
When only superscripts are used, they behave in exactly the same manner as when only subscripts are used.

{\footnotesize
\begin{verbatim}
[|[|11 12 13|] [|21 22 23|] [|31 32 33|]|]~1~1
;11

[|[|[|1 2|] [|3 4|]|] [|[|5 6|] [|7 8|]|]|]~i~j~i
;[|[|1 3|] [|6 8|]|]~i~j
\end{verbatim}
}

The index reduction rules thus far are the same as those of the existing work~\cite{aahlander2002einstein}.

Let us consider a case in which the same symbols are used for a superscript and a subscript.
In this case, the tensor is automatically contracted using ``$+$'' in the existing research.
In contrast, Egison converts it to the tensor composed of diagonal components, as in the above examples.
However, in that case, the summarized indices become a \textit{supersubscript}, which is represented by ``\verb|~_|''.

{\footnotesize
\begin{verbatim}
[|[|11 12 13|] [|21 22 23|] [|31 32 33|]|]~i_i
;[|11 22 33|]~_i
\end{verbatim}
}

Even when three or more indices of the same symbol appear that contain both supersubscripts and subscripts, Egison converts it to the tensor composed of diagonal components for all these indices.

{\footnotesize
\begin{verbatim}
[|[|[|1 2|] [|3 4|]|] [|[|5 6|] [|7 8|]|]|]~i~i_i
;[|1 8|]~_i
\end{verbatim}
}

The reason not to contract it immediately is to enable it to parameterize an operator for contraction.
The components of supersubscripts can be contracted by using the \texttt{contract} expression.
The \texttt{contract} expression receives a function to be used for contraction as the first argument, and a target tensor as the second argument.

{\footnotesize
\begin{verbatim}
(contract + [|11 22 33|]~_i)
;66
\end{verbatim}
}

\begin{figure}
  \begin{center}
{\footnotesize
\begin{verbatim}
E({A, xs}) =
  if e(xs) = [] then
    {A, xs})
  elsif e(xs) = [{k,j}, …] & p(k,xs) = p(j,xs) then
    E({diag(k, j, A), remove(j, xs))
  elsif e(xs) = [{k,j}, …] & p(k,xs) != p(j,xs) then
    E({diag(k, j, A), update(k, 0, remove(j, xs)))
\end{verbatim}
}
  \end{center}
  \caption{Pseudo code of index reduction}
  \label{fig:semanticsRules}
\end{figure}

Figure~\ref{fig:semanticsRules} shows pseudo code of index reduction as explained in this section.
\texttt{E(A,xs)} is a function for reducing a tensor with indices.
\texttt{A} is an array that consists of tensor components.
\texttt{xs} is a list of indices appended to \texttt{A}.
For example, \texttt{E(A,xs)} works as follows with the tensor whose indices are ``\verb|~i_j_i|''.
We use ``\texttt{1}'', ``\texttt{-1}'', and ``\texttt{0}'' to represent a superscript, subscript, and supersubscript, respectively.

{\footnotesize
\begin{verbatim}
E({[|[|[|1 2|] [|3 4|]|] [|[|5 6|] [|7 8|]|]|],
  [{i,1}, {j,1}, {i,-1}]}) =
{[|[|1 3|] [|6 8|]|], [{i,0}, {j,1}]}
\end{verbatim}
}

Next, we explain the helper functions used in Figure~\ref{fig:semanticsRules}.
\texttt{e(xs)} is a function for finding pairs of identical indices from \texttt{xs}.
\texttt{diag(k, j, A)} is a function for creating the tensor that consists of diagonal components of \texttt{A} for the \texttt{k}-th and \texttt{j}-th order.
\texttt{remove(k, xs)} is a function for removing the \texttt{k}-th element from \texttt{xs}.
\texttt{p(k, xs)} is a function for obtaining the value of the \texttt{k}-th element of the assoc list \texttt{xs}.
\texttt{update(xs, k, p)} is a function for updating the value of the \texttt{k}-th element of the assoc list \texttt{xs} to \texttt{p}.
These functions work as follows.

{\footnotesize
\begin{verbatim}
e([{i, 1}, {j, -1}, {i, 1}])        = [{1,3}]
diag(1, 2, [|[|11 12|] [|21 22|]|]) = [|11 22|]
p(2, 0, [{i, 1}, {j, -1}])          = -1
remove(2, [{i, 1}, {j, -1}])        = [{i, 1}]
update(2, 0, [{i, 1}, {j, -1}])     = [{i, 1}, {j, 0}]
\end{verbatim}
}

\subsection{Scalar and Tensor Functions}\label{sfTf}

In this section we introduce the concept of two types of functions: scalar and tensor functions.
This enables us to introduce naturally the parameterization of an operator for tensor product and contraction when we combine it with the index reduction rules explained in Section~\ref{egisonIndex}.

Tensor functions are functions that involve contraction of the tensors provided as an argument, and scalar functions are not.
For example, the inner product of vectors and matrix multiplication are tensor functions.
In contrast, ``\texttt{+}'', `` \texttt{-}'', ``\texttt{*}'', and ``\texttt{/}'' are scalar functions.

We use the concept of two types of parameters, \textit{scalar parameters} and \textit{tensor parameters}, to specify whether the function we defined is a scalar or tensor function.
Similar to Scheme, Egison generates a function using a \texttt{lambda} expression.
In the \texttt{lambda} expression, we add ``\texttt{\$}'' or ``\texttt{\%}'' to the beginning of the parameters.
A parameter to which ``\texttt{\$}'' is prefixed is a scalar parameter.
A parameter to which ``\texttt{\%}'' is prefixed is a tensor parameter.

As with ordinary parameters, when a tensor parameter obtains a tensor as an argument, the function treats the tensor as it is.
In contrast, when a scalar parameter obtains a tensor as an argument, the function is applied to each component of the tensor.
A function with scalar parameters is converted to a function only with tensor parameters by using the \texttt{tensor-map} function as follows.
In this way, we can implement scalar parameters.

{\footnotesize
\begin{verbatim}
(lambda [$x $y] ...)
;=>(lambda [%x %y]
     (tensor-map (lambda [%x]
                   (tensor-map (lambda [%y] ...)
                               y))
                 x))
\end{verbatim}
}

As the name implies, the \texttt{tensor-map} function applies the function of the first argument to each component of the tensor provided as the second argument.
If the result of applying the function of the first argument to each component of the tensor provided as the second argument is a tensor with indices, it moves those indices to the end of the tensor that is the result of evaluating the \texttt{tensor-map} function.
We will see an example in a later part of this section.

Let's review the \texttt{min} function defined in Figure~\ref{fig:minFunction} in Section~\ref{intro} as an example of a scalar function.
This \texttt{min} function can handle tensors as arguments as follows.

{\footnotesize
\begin{verbatim}
(min [|1 2 3|]_i [|10 20 30|]_j)
;[|[|1 1 1|] [|2 2 2|] [|3 3 3|]|]_i_j

(min [|1 2 3|]_i [|10 20 30|]_i)
;[| 1 2 3 |]_i;
\end{verbatim}
}

Note that the tensor indices of the evaluated result are ``\verb|_i_j|''.
If the \texttt{tensor-map} function simply applies the function to each component of the tensor, the result of this program will be similar to \\
``\verb![|[|1 1 1|]_j [|2 2 2|]_j [|3 3 3|]_j|]_i!''.
However, as explained above if the results of applying the function to each component of the tensor are tensors with indices, it moves those indices to the end of the tensor that is the result of evaluating the \texttt{tensor-map} function.
This is the reason that the indices of the evaluated result are ``\verb|_i_j|''.
This mechanism enables us to directly apply scalar functions to tensor arguments using index notation as the above example.

The above evaluation result is equal to the result of specifying the \texttt{min} function as the operator of the tensor product.
By defining a scalar function as described above, the parameterization of an operator for the tensor product is achieved without bringing it to programmers' attention.
Proposing the concept of scalar functions is one of the major contributions of this paper.

Next, let's review the ``\texttt{.}'' function defined in Figure~\ref{fig:dotFunction} in Section~\ref{intro} as an example of a tensor function.
All of the parameters of the ``\texttt{.}'' function are tensor parameters and ``\texttt{\%}'' is prepended to the beginning of the parameters.
This ``\texttt{.}'' function can handle tensors as arguments as follows.

{\footnotesize
\begin{verbatim}
(. [|1 2 3|]~i [|10 20 30|]_i)
;140

(. [|1 2 3|]_i [|10 20 30|]_j)
;[| [| 10 20 30 |] [| 20 40 60 |] [| 30 60 90 |] |]_i_j

(. [|1 2 3|]_i [|10 20 30|]_i)
;[| 10 40 90 |]_i
\end{verbatim}
}

When a tensor with indices is given as an argument of a tensor function, it is passed to the tensor function maintaining its indices.
Note that we can directly apply tensor functions to tensor arguments using index notation as in the above example.

By changing ``\texttt{*}'' and ``\texttt{+}'' appearing in Figure~\ref{fig:dotFunction} to different functions, we can define a new tensor multiplication operator that uses the functions we specified to calculate the tensor product and the sum of the traces for contraction.
  
Since a tensor parameter is used only when defining a function that contracts tensors, in most cases only scalar parameters are used.

\subsection{Application of Scalar Functions to Tensors}\label{egisonOp}

In Section~\ref{sfTf}, we introduced scalar and tensor parameters, and explained the behavior of scalar and tensor functions.
The definition of scalar functions in Section~\ref{sfTf} is highly compatible with the index reduction rules for a single tensor with indices explained in Section~\ref{egisonIndex}.
In this section, we confirm this fact seeing various samples of a scalar function receiving indexed tensors as arguments.

In the following sample, the tensor that has the subscript ``\texttt{i}'' and the subscript ``\texttt{j}'' is applied to the scalar function ``\texttt{+}''.
If the indices of the tensors of the arguments are different in this manner, it returns the tensor product using the scalar function as the operator, as we saw in Section ~\ref{sfTf}.

{\footnotesize
\begin{verbatim}
(+ [|1 2 3|]_i [|10 20 30|]_j)
;[|[|11 21 31|] [|12 22 32|] [|13 23 33|]|]_i_j
\end{verbatim}
}

In the following example, we add the same index ``\texttt{i}'' to both \texttt{[|1 2 3|]} and \texttt{[|10 20 30|]}.
When the indices of the tensors given as arguments are identical, the scalar function is applied to each corresponding component.
An error occurs if the dimensions are different even though the indices are identical.
Note that this result is equal to the result of simplifying ``\verb![|[|11 21 31|] [|12 22 32|] [|13 23 33|]|]_i_i!'' by the reduction rules in Section~\ref{egisonIndex}.
Review Figure~\ref{fig:inPlusIndex} to clarify the idea underlying this transformation.

{\footnotesize
\begin{verbatim}
(+ [|1 2 3|]_i [|10 20 30|]_i)
;[|11 22 33|]_i
\end{verbatim}
}

Both arguments are vectors in the above two examples.
Next, let us see examples in which the arguments are higher-order tensors, as follows.
We can see that the reduction rules work well even for high-order tensors.

{\footnotesize
\begin{verbatim}
(+ [|[|11 12|] [|21 22|] [|31 32|]|]_i_j [|100 200 300|]_i)
;[|[|111 112|] [|221 222|] [|331 332|]|]_i_j

(+ [|[|1 2 3|] [|10 20 30|]|]_i_j [|100 200 300|]_j)
;[|[|101 202 303|] [|110 220 330|]|]_i_j
\end{verbatim}
}

As mentioned in Section~\ref{sfTf}, arbitrary scalar functions behave in the same manner as the above examples.

The ``\texttt{∂/∂}`' function appearing in Figure~\ref{fig:inEgisonShort} is also a scalar function.
However, ``\texttt{∂/∂}'' is not a normal scalar function.
``\texttt{∂/∂}''  is a scalar function that inverts indices of the tensor given as its second argument.
For example, the program ``\verb|(∂/∂ Γ~i_j_k x~l)|'' returns the fourth-order tensor with superscript ``\texttt{i}'', subscript ``\texttt{j}'', subscript ``\texttt{k}'', and subscript ``\texttt{l}'' from left to right.

To define scalar functions such as ``\texttt{∂/∂}'', we use \textit{inverted scalar parameters}.
Inverted scalar parameters are represented by ``\texttt{*\$}''.
A program that uses inverted scalar parameters is transformed as follows.
Here, the \texttt{flip-indices} function is a primitive function for inverting the indices of a tensor provided as an argument upside down.
Supersubscripts remain as supersubscripts even if they are inverted.

{\footnotesize
\begin{verbatim}
(define $∂/∂ (lambda [$f *$x] ...))
;=>(define $∂/∂ (lambda [%f %x]
     (tensor-map (lambda [%x] (tensor-map (lambda [%y] ...)
                                          (flip-indices x)))
                 f))
\end{verbatim}
}
    
The definition of ``\texttt{∂/∂}'' can be seen in the GitHub repository\footnote{\url{https://github.com/egison/egison/blob/master/lib/math/analysis/derivative.egi}}.

In the following example, we apply ``\texttt{∂/∂}'' to tensors.

{\footnotesize
\begin{verbatim}
(∂/∂ [|(* r (sin θ)) (* r (cos θ))|]_i [|r θ|]_j)
;[|[|(sin θ) (* r (cos θ))|]
;  [|(cos θ) (* -1 r (sin θ))|]|]_i~j

(∂/∂ [|(* r (sin θ)) (* r (cos θ))|]_i [|r θ|]_i)
;[|(sin θ) (* -1 r (sin θ))|]~_i
\end{verbatim}
}

In the writing of a program that deals with high-order tensors, the number of symbols used for indices increases.
A dummy symbol is introduced to suppress that.
``\texttt{\#}'' represents a dummy symbol.
All instances of ``\texttt{\#}'' are treated as different symbols.
Using this mechanism makes it easier to distinguish indices that are important in the program, thereby also improving the readability of the program.

The idea of dummy symbols is not new.
For example, there is syntax to generate local symbols in the Wolfram languages~\cite{wolframModule}~ and Maxima\cite{sympyDummySymbol}.
We can use such syntax to generate dummy symbols, though its primary purpose is to generate temporary symbols for substituting variables.

The novelty of this paper on dummy symbols is that we prepared one-character syntax ``\texttt{\#}'' to generate dummy symbols to be used as indices of tensors.
This is a very simple idea, but since this notation is not used even in mathematics, we think this idea is new.

{\footnotesize
\begin{verbatim}
(+ [|1 2 3|]_# [|10 20 30|]_#)
;[|[|11 21 31|] [|12 22 32|] [|13 23 33|]|]_#_#
\end{verbatim}
}

Egison allows programmers to omit indices while recommending that programmers explicitly specify indices.
If the indices are omitted, Egison handles the expression in the same manner as dummy symbols are omitted.

{\footnotesize
\begin{verbatim}
(+ [|1 2 3|] [|10 20 30|])
;[|[|11 21 31|] [|12 22 32|] [|13 23 33|]|]
\end{verbatim}
}

Other computer algebra systems, including the Wolfram language, handle expressions in the same manner, as the same indices are added to all arguments where indices are omitted.

{\footnotesize
\begin{verbatim}
[10,20,30] + [1,2,3] = [11,22,33]
\end{verbatim}
}

Egison selects the above specification to avoid creating difficulty for the interpreter in complementing the part of an expression in which programmers omit the description.
Regardless of which specification is adopted, when a scalar function takes a scalar and a tensor as arguments, it performs as follows.

{\footnotesize
\begin{verbatim}
(+ [|1 2 3|] 10)
;[|11 12 13|]
\end{verbatim}
}

\subsection{The \texttt{with-symbols} Expression}\label{withSymbols}

The \texttt{with-symbols} expression is syntax for generating new local symbols, such as the \texttt{Module} expression in the Wolfram language~\cite{wolframModule}.

One-character symbols that are often used as indices of tensors such as ``\texttt{i}'', ``\texttt{j}'', and ``\texttt{k}'' are often used in another part of a program.
Generating local symbols using \texttt{with-symbols} expressions enables us to avoid variable conflicts for such symbols.

The \texttt{with-symbols} expression takes a list of symbols as its first argument.
These symbols are valid only in the expression given in the second argument of the \texttt{with-symbols} expression.

{\footnotesize
\begin{verbatim}
(with-symbols {i} (contract + (* [|1 2 3|]~i [|10 20 30|]_i)))
;60
\end{verbatim}
}

If the evaluation result of the body of the \texttt{with-symbols} expression contains the symbols generated by the \texttt{with-symbols} expressions, those symbols are converted into the dummy symbols described in Section~\ref{egisonOp}.
As a result of this mechanism, local symbols never appear in the result of the \texttt{with-symbols} expression.
This mechanism enables us to use local symbols as indices of tensors that often remain in the evaluation result.

{\footnotesize
\begin{verbatim}
(with-symbols {i} (+ [|1 2 3|]_i [|10 20 30|]_i))
;[|11 22 33|]_#
\end{verbatim}
}

\subsection{Definitions of Tensor Functions}\label{egisonPara}

In this section, we define the function that calculates the inner product of vectors and matrix multiplication.

We can define the function for calculating the inner product as follows.
This function is a tensor function, because it contracts the tensor.

{\footnotesize
\begin{verbatim}
(define $inner-product (lambda [%v1 %v2]
    (with-symbols {i} (contract + (* v1~i v2_i)))))
\end{verbatim}
}

Unlike the ``\texttt{.}'' function defined in Section~\ref{sfTf}, this \texttt{inner-product} function assumes that no index is appended to the vectors given as arguments, such as ``\texttt{(inner-product [|1 2 3|] [|10 20 30|])}''.
Therefore, the index is appended in the function definition as ``\verb|v1~i|'' and ``\verb|v2_i|''.
If the vector provided as an argument has an index, it will be overwritten.

We can define the function for multiplying matrices as follows.
This function is a tensor function as well.

{\footnotesize
\begin{verbatim}
(define $mat-mul (lambda [%m1 %m2]
    (with-symbols {j} (contract + (* m1~#~j m2_j_#)))))
\end{verbatim}
}

The \texttt{mat-mul} function also assumes that no indices are appended to the matrices provided as arguments.
Therefore, the indices are appended in the function definition as ``\verb|m1~#_j|'' and ``\verb|m2_j_#|''.
Note that dummy symbols are again used effectively in this program.
Without dummy symbols, the program ``\verb|(* m1~#~j m2_j_#)|'' is represented ``\verb|(* m1~i~j m2_j_k)|''.
We avoid declaring the additional symbols ``\texttt{i}'' and ``\texttt{k}'' by using dummy symbols.

Thus, if we define a tensor function using \texttt{with-symbols} expressions, we can define functions that can be used without knowledge of index notation.

%
%
%
%
%
%

\subsection{Tensor Generation Syntax and Pattern-Matching}

Egison has a \texttt{generate-tensor} expression that is tensor generation syntax having essentially the same meaning as the \texttt{Table} expression in the Wolfram language.
This syntax is even more powerful and provides for programs that initiate a complicated matrix more simply when combined with nonlinear pattern-matching implemented in Egison~\cite{egi2014non}.
This section discusses this feature.

The \texttt{generate-tensor} expression takes a function as the first argument and the size of the tensor to be generated as the second argument.
The number of arguments of the function of the first argument is equal to the rank of the tensor to be generated.
Each component of the generated tensor is the result of applying the indices to the function of the first argument.

For example, a unit matrix can be initialized as follows.

{\footnotesize
\begin{verbatim}
(generate-tensor
  (match-lambda [integer integer]
    {[[$i ,i] 1] [[_ _] 0]})
  {4 4})
;[|[|1 0 0 0|] [|0 1 0 0|] [|0 0 1 0|] [|0 0 0 1|]|]
\end{verbatim}
}

As a more complicated example, the matrix for generating the N-bonacci sequence can be initialized as follows.
A program for calculating N-bonacci sequence using this matrix is available in the Egison website\footnote{\url{https://www.egison.org/math/tribonacci.html}}.

{\footnotesize
\begin{verbatim}
(generate-tensor
  (match-lambda [integer integer]
    {[[,1 _] 1]
     [[$x ,(- x 1)] 1]
     [[_ _] 0]})
  {4 4})
;[|[|1 1 1 1|] [|1 0 0 0|] [|0 1 0 0|] [|0 0 1 0|]|]
\end{verbatim}
}

We consider the fact that pattern-matching is useful for initializing complicated tensors in a simple manner to be evidence for the necessity of pattern-matching with strong expressive power, as discussed in~\cite{egi2014non}.
We will not discuss this in more detail here, because it is not the principal topic of this paper.

\section{Calculation of Riemann Curvature Tensor}\label{demo}

In this section, we present programs for calculating the Riemannian curvature tensor~\cite{fleisch2011student,schutz1980geometrical,ollivier2011visual}, the fourth-order tensor that expresses the curvature of a manifold, in the Wolfram language and Egison, to shows the advantages of our proposal.

Figures~\ref{fig:inWolfram} and \ref{fig:inEgison} show programs for calculating the Riemann curvature tensor of a torus in the Wolfram language and in Egison, respectively.

In Egison, when binding a tensor to a variable, we can specify the type of indices in the variable name.
For example, we can bind different tensors to ``\verb|$g__|'', ``\verb|$g~~|'', ``\verb|$Γ___|'', and ``\verb|Γ~__|''.
This feature is also implemented in the existing work described in Section~\ref{maxima}.
This feature simplifies variable names.

In Egison, some of the tensors are bound to a variable with symbolic indices such as ``\verb|$Γ_i_j_k|''.
It is automatically desugared as follows.
This syntactic sugar renders a program closer to the mathematical expression.
\texttt{transpose} is a function for transposing the tensor in the second argument as specified in the first argument.

{\footnotesize
\begin{verbatim}
(define $Γ_i_j_k ...)
;=>(define $Γ___
     (with-symbols {i j k} (transpose {i j k} ...)))
\end{verbatim}
}

Except for that point, the programs for calculating the torus coordinates and the metric tensor differ only in the appearance of the syntax.
On the other hand, our proposal introduces essential differences in the programs for calculating the local basis, Christoffel symbols of the first and second kind, and Riemann curvature tensor.
This section explains these differences.

First of all, the Wolfram language uses the \texttt{Table} expression in the program for calculating the local basis, but Egison has a flat description using scalar functions.
The \texttt{flip} function used in the Egison program is a function for swapping the arguments of a two-argument function.
It is used to transpose the result matrix.

Next, let us examine the programs for calculating Christoffel symbols of the first and second kind, and the Riemann curvature tensor.
They are defined in mathematics as follows.

\[\Gamma_{ijk} = \frac{1}{2} (\frac{\partial g_{ij}}{\partial x^k} + \frac{\partial g_{ik}}{\partial x^j} - \frac{\partial g_{kj}}{\partial x^i})\]

\[\Gamma^{i}_{\;kl} = g^{ij} \Gamma_{jkl}\]

\[R^{i}_{\;jkl} = \frac{\partial \Gamma^{i}_{\;jl}}{\partial x^k} - \frac{\partial \Gamma^{i}_{\;jk}}{\partial x^l} + \Gamma^{m}_{\;jl} \Gamma^{i}_{\;mk} - \Gamma^{m}_{\;jk} \Gamma^{i}_{\;ml} \]

The essential difference between Figures~\ref{fig:inWolfram} and ~\ref{fig:inEgison} is that the Wolfram language uses \texttt{Table} expressions to represent the above three equations, whereas Egison uses index notation directly.
In particular, in the program for calculating the Riemann curvature tensor, a double loop consisting of the \texttt{Table} and \texttt{Sum} expressions appears in the Wolfram language, whereas the Egison program is as flat as the mathematical expression.

From the above discussion, we can conclude that Egison expresses mathematical expressions more directly than the Wolfram language, though there is little difference in the number of lines in the programs.

\begin{figure}[t]
  \begin{center}
{\footnotesize
\begin{verbatim}
(* Coordinates for Torus *)
M=2;
x={θ,φ};
X={(a*Cos[θ]+b)Cos[φ], (a*Cos[θ]+b)Sin[φ], a*Sin[θ]};

(* Local basis *)
e=Table[D[X[[j]],x[[i]]],{i,2},{j,3}] //ExpandAll//Simplify;

(* Metric tensor *)
g=Table[e[[i]].e[[j]],{i,M},{j,M}] //ExpandAll//Simplify;
Ig=Inverse[g] //ExpandAll//Simplify;

(* Christoffel symbols of the first kind *)
Γ1=Table[D[g[[i,j]],x[[k]]] + D[g[[i,k]],x[[j]]]
         - D[g[[j,k]],x[[i]]],
         {i,M},{j,M},{k,M}]/2 //ExpandAll//Simplify;

(* Christoffel symbols of the second kind *)
Γ2=Table[Sum[Ig[[i,j]]Γ1[[j,k,l]],{j,M}],
         {i,M},{k,M},{l,M}] //ExpandAll//Simplify;

(* Riemann curvature tensor *)
R=Table[D[Γ2[[i,j,l]],x[[k]]] - D[Γ2[[i,j,k]],x[[l]]]
       +Sum[Γ2[[m,j,l]]Γ2[[i,m,k]]
          - Γ2[[m,j,k]]Γ2[[i,m,l]],
            {m,M}],
        {i,M},{j,M},{k,M},{l,M}] //ExpandAll//Simplify;
\end{verbatim}
}
  \end{center}
  \caption{A program to calculate Riemann curvature tensor in Wolfram language}
  \label{fig:inWolfram}
\end{figure}
\begin{figure}[t]
  \begin{center}
{\footnotesize
\begin{verbatim}
;; Coordinates for Torus
(define $x [|θ φ|])
(define $X [|(* '(+ (* a (cos θ)) b) (cos φ)) ; = x
             (* '(+ (* a (cos θ)) b) (sin φ)) ; = y
             (* a (sin θ))|])                 ; = z

;; Local basis
(define $e ((flip ∂/∂) x~# X_#))

;; Metric tensor
(define $g__ (generate-tensor 2#(V.* e_%1 e_%2) {2 2}))
(define $g~~ (M.inverse g_#_#))

;; Christoffel symbols of the first kind
(define $Γ_i_j_k
  (* (/ 1 2)
     (+ (∂/∂ g_i_j x_k)
        (∂/∂ g_i_k x_j)
        (* -1 (∂/∂ g_j_k x_i)))))

;; Christoffel symbols of the second kind
(define $Γ~__ (with-symbols {i} (. g~#~i Γ_i_#_#)))

;; Riemann curvature tensor
(define $R~i_j_k_l
  (with-symbols {m}
    (+ (- (∂/∂ Γ~i_j_l x_k) (∂/∂ Γ~i_j_k x_l))
       (- (. Γ~m_j_l Γ~i_m_k) (. Γ~m_j_k Γ~i_m_l)))))
\end{verbatim}
}
  \end{center}
  \caption{A program to calculate Riemann curvature tensor in Egison}
  \label{fig:inEgison}
\end{figure}

\section{Conclusion}\label{conclusion}

\subsection{Contributions}

In this paper, we introduced index notation with simpler index rules than in the existing work, as explained in Sections~\ref{egisonIndex} and~\ref{sfTf}.
Additionally, we introduced the concept of two types of functions, scalar and tensor functions, as explained in Sections~\ref{sfTf} and~\ref{egisonOp}.
We showed that the combination of these two ideas enables us to directly apply arbitrary user-defined functions to tensor arguments using index notation.
We also proposed that these two types of functions can be defined using two types of parameters, scalar and tensor parameters.
This proposal eliminates the need to consider the case in which the argument is a tensor when defining scalar functions, which appear in an overwhelming proportion of programs.

In addition, our proposal achieved lexical scoping of symbols used as tensor indices by the \texttt{with-symbols} expression and a dummy symbol ``\texttt{\#}'', as explained in Sections~\ref{withSymbols} and~\ref{egisonPara}.
This is also our important contribution since there is no literature that discuss this topic.

This paper also showed the usefulness of a dummy symbol ``\texttt{\#}'' and of using it as an index of tensors.
This idea allows us to reduce the number of symbols used as indices by replacing symbols that appear only once with dummy symbols ``\texttt{\#}''.
Although this is a very simple idea, it improves the readability of programs by highlighting important indices.

\subsection{Future Work}

In this paper, we introduced several forms of syntax into a language to introduce the new concepts of scalar and tensor functions.
However, we think it is possible to introduce the concepts of scalar and tensor functions even using a static type system or the overloading feature of object-oriented programming.
For example, in a static type system, whether the parameter of a function is a scalar or tensor parameter can be specified when specifying the type of the function.
Although we could not discuss this in this paper, it is an interesting research topic to think about how to incorporate the ideas proposed in this paper into existing programming languages naturally.

In particular, it is of substantial significance to incorporate this method into programming languages such as Formura~\cite{muranushi2016automatic} and Diderot~\cite{kindlmann2016diderot} that have a compiler that generates code for executing tensor calculation in parallel.
For example, incorporating this method into Formura would enable us to describe physical simulation using not only the Cartesian coordinate system but also more general coordinate systems such as the polar and spherical coordinate system in simple programs.

By the way, index notation as discussed in this paper is a notation invented over a century ago.
Especially, it is well known that Einstein summation notation was invented by Einstein when he was working on general relativity theory.
In addition to index notation, there might still be many notations in mathematics that are useful, but not yet introduced into programming.
There might also be notations that describe the formulas of existing theories more concisely, but that mathematicians have not discovered yet.

We contend that it is very useful for those researching programming languages who are familiar with many programming paradigms and can flexibly create new programming languages to learn a wider range of mathematics for the future of programming languages.

\section*{Acknowledgement}
I thank Hiromi Hirano, Hidehiko Masuhara, and Michal J. Gajda for very helpful feedback on the earlier versions of the paper.

%

%


\bibliographystyle{abbrvnat}


\bibliography{tegison-arxiv}

\begin{thebibliography}{17}
\providecommand{\natexlab}[1]{#1}
\providecommand{\url}[1]{\texttt{#1}}
\expandafter\ifx\csname urlstyle\endcsname\relax
  \providecommand{\doi}[1]{doi: #1}\else
  \providecommand{\doi}{doi: \begingroup \urlstyle{rm}\Url}\fi

\bibitem[max(2016)]{maximaWeb}
{Maxima - a Computer Algebra System}, 2016.
\newblock \url{http://maxima.sourceforge.net/}.

\bibitem[sym(2016)]{sympyDummySymbol}
{SymPy User’s Guide - SymPy 1.0.1.dev documentation}, 2016.
\newblock \url{http://docs.sympy.org/dev/guide.html}.

\bibitem[{\AA}hlander(2002)]{aahlander2002einstein}
K.~{\AA}hlander.
\newblock Einstein summation for multidimensional arrays.
\newblock \emph{Computers \& Mathematics with Applications}, 44\penalty0
  (8-9):\penalty0 1007--1017, 2002.

\bibitem[Egi(2014)]{egi2014non}
S.~Egi.
\newblock Non-linear pattern-matching against non-free data types with lexical
  scoping.
\newblock \emph{arXiv preprint arXiv:1407.0729}, 2014.

\bibitem[Egi(2016)]{egison}
S.~Egi.
\newblock {The Egison Programming Language}, 2016.
\newblock \url{https://www.egison.org}.

\bibitem[Fleisch(2011)]{fleisch2011student}
D.~A. Fleisch.
\newblock \emph{A student's guide to vectors and tensors}.
\newblock Cambridge University Press, 2011.

\bibitem[Hartley and Zisserman(2003)]{hartley2003multiple}
R.~Hartley and A.~Zisserman.
\newblock \emph{Multiple view geometry in computer vision}.
\newblock Cambridge university press, 2003.

\bibitem[Kindlmann et~al.(2016)]{kindlmann2016diderot}
G.~Kindlmann et~al.
\newblock Diderot: a domain-specific language for portable parallel scientific
  visualization and image analysis.
\newblock \emph{IEEE transactions on visualization and computer graphics},
  22\penalty0 (1):\penalty0 867--876, 2016.

\bibitem[Maeda et~al.(2010)]{maeda2010program}
Y.~Maeda et~al.
\newblock {Computation of the Wodzicki-Chern-Simons form in local coordinates.
  Computations for $S^1$ actions on $S^2 \times S^3$}, 2010.
\newblock
  \url{http://math.bu.edu/people/sr/articles/ComputationsChernSimonsS2xS3_July_1_2010.pdf}.

\bibitem[Maeda et~al.(2016)]{maeda2016geometry}
Y.~Maeda et~al.
\newblock The geometry of loop spaces ii: Characteristic classes.
\newblock \emph{Advances in Mathematics}, 287:\penalty0 485--518, 2016.

\bibitem[Muranushi et~al.(2016)]{muranushi2016automatic}
T.~Muranushi et~al.
\newblock Automatic generation of efficient codes from mathematical
  descriptions of stencil computation.
\newblock In \emph{Proceedings of the 5th International Workshop on Functional
  High-Performance Computing}, pages 17--22. ACM, 2016.

\bibitem[Ollivier(2011)]{ollivier2011visual}
Y.~Ollivier.
\newblock A visual introduction to riemannian curvatures and some discrete
  generalizations.
\newblock \emph{Analysis and Geometry of Metric Measure Spaces: Lecture Notes
  of the 50th S{\'e}minaire de Math{\'e}matiques Sup{\'e}rieures (SMS),
  Montr{\'e}al}, pages 197--219, 2011.

\bibitem[Schutz(1980)]{schutz1980geometrical}
B.~F. Schutz.
\newblock \emph{Geometrical methods of mathematical physics}.
\newblock Cambridge university press, 1980.

\bibitem[Solomonik and Hoefler(2015)]{solomonik2015sparse}
E.~Solomonik and T.~Hoefler.
\newblock Sparse tensor algebra as a parallel programming model.
\newblock \emph{arXiv preprint arXiv:1512.00066}, 2015.

\bibitem[Toth(2005)]{toth2005tensor}
V.~Toth.
\newblock Tensor manipulation in gpl maxima.
\newblock \emph{arXiv preprint cs/0503073}, 2005.

\bibitem[Wolfram(2016{\natexlab{a}})]{wolframModule}
Wolfram.
\newblock {Module - Wolfram Language Documentation}, 2016{\natexlab{a}}.
\newblock \url{http://reference.wolfram.com/language/ref/Module.html}.

\bibitem[Wolfram(2016{\natexlab{b}})]{wolframTable}
Wolfram.
\newblock {Table - Wolfram Language Documentation}, 2016{\natexlab{b}}.
\newblock \url{http://reference.wolfram.com/language/ref/Table.html}.

\end{thebibliography}


\end{document}